\documentclass[acmsmall,screen,table]{acmart}

\usepackage{tabularx,booktabs,ragged2e,siunitx}
\newcolumntype{C}{>{\Centering\arraybackslash}X}
\newcolumntype{L}{>{\RaggedRight\arraybackslash\hspace{0pt}}X}

\usepackage{subcaption}
\usepackage{multirow}
\makeatletter
\def\namedlabel#1#2{\begingroup
    #2%
    \def\@currentlabel{#2}%
    \phantomsection\label{#1}\endgroup
}
\makeatother

\usepackage{graphicx}
\usepackage{booktabs}
\usepackage{layouts}  
\usepackage{subcaption} 
\usepackage{makecell} 
\usepackage{multirow} 
\usepackage[normalem]{ulem} 
\useunder{\uline}{\ul}{} 
\usepackage{tabularx}
\usepackage{subcaption}

\theoremstyle{definition}

\definecolor{offwhite}{gray}{0.92}  
\definecolor{mygreen}{HTML}{91cf60}

\usepackage{csquotes}
\usepackage[normalem]{ulem}

\usepackage{xcolor}

\AtBeginDocument{%
  }

\begin{document}
\title{Beyond the Individual: A Community-Engaged Framework for Ethical Online Community Research}




\author{Matthew Zent}
\email{zentx005@umn.edu}
\orcid{0000-0003-4555-8764}
\affiliation{%
  \institution{University of Minnesota}
  \country{USA}
}
\author{Seraphina Yong}
\email{yong0021@umn.edu}
\orcid{0000-0002-9359-2302}
\affiliation{%
  \institution{University of Minnesota}
  \country{USA}
}
\author{Dhruv Bala}
\email{balas095@alumni.umn.edu}
\orcid{0009-0005-9179-8752}
\affiliation{%
  \institution{University of Minnesota}
  \country{USA}
}
\author{Stevie Chancellor}
\email{steviec@umn.edu}
\orcid{0000-0003-0620-0903}
\affiliation{%
  \institution{University of Minnesota}
  \country{USA}
}
\author{Joseph A. Konstan}
\email{Konstan@umn.edu}
\orcid{0000-0002-7788-2748}
\affiliation{%
  \institution{University of Minnesota}
  \country{USA}
}
\author{Loren Terveen}
\email{terveen@umn.edu}
\orcid{0000-0002-8843-4035}
\affiliation{%
  \institution{University of Minnesota}
  \country{USA}
}
\author{Svetlana Yarosh}
\email{lana@umn.edu}
\orcid{0000-0001-8389-2064}
\affiliation{%
  \institution{University of Minnesota}
  \country{USA}
}

\renewcommand{\shortauthors}{Zent et al.}

\begin{abstract}
  Online community research routinely poses minimal risk to individuals, but does the same hold true for online communities? In response to high-profile breaches of online community trust and increased debate in the social computing research community on the ethics of online community research, this paper investigates community-level harms and benefits of research. Through 9 participatory-inspired workshops with four critical online communities (Wikipedia, InTheRooms, CaringBridge, and r/AskHistorians) we found researchers should engage more directly with communities' primary purpose by rationalizing their methods and contributions in the context of community goals to equalize the beneficiaries of community research. To facilitate deeper alignment of these expectations, we present the FACTORS (Functions for Action with Communities: Teaching, Overseeing, Reciprocating, and Sustaining) framework for ethical online community research. Finally, we reflect on our findings by providing implications for researchers and online communities to identify and implement functions for navigating community-level harms and benefits.  
\end{abstract}


\maketitle
\section{Introduction}

Online communities have long been a central focus of Human-Computer Interaction (HCI) and Computer-Supported Cooperative Work (CSCW) research~\cite{gatos_how, malinen2015understanding, harris_joining}.
This extended history studied online groups to understand many sociotechnical topics, including why members participate~\cite{panciera_wikipedians, hwang_why}, rules and norms~\cite{chandrasekharan_internet}, health support~\cite{smith_cannot, papoutsaki_future}, and the cooperative work these communities perform~\cite{wallace_technologies}.
As a result, the social computing research community has had ample debates about the ethics of internet studies~\cite{densmore_research, fiesler_sigchi, fiesler_hci, fiesler_internet}.



Despite the prevalence of online community research, there are no well-established ethical principles for doing research that may impact an online community~\cite{kantanen_hazy_2016, sugiura_ethical_2017}. 
There are entire fields of research devoted to ethical research on individuals~\cite{farrimond2012doing, moor2020computer} as well as interest in HCI/CSCW~\cite{oppermann_beyond, rifat2023many, fleischmann_good}. However, these guidelines for research on individuals may be significantly different from the expectations of the members of communities (e.g.,~\cite{fiesler_participant_2018}). 
High-profile ethical breaches in news headlines provide evidence of this misalignment.
For instance, the Linux Hypocrite Commits~\cite{clark_university_2021} and Princeton-Radboud Study on Privacy Law Implementation~\cite{jonathan_mayer_note_2021} demonstrated a mismatch between Institutional Review Board (IRB) and community interpretations of human-subjects research leading to severed community-researcher trust.
Similarly, the Facebook Emotional Contagion study~\cite{hallinan_unexpected_2020} revealed a lack of shared understanding within the community itself regarding the ethical implications of platform-based research, which negatively impacted community health as individual harms led to member attrition.
These cases leave online communities reeling and research institutions subjected to collective shaming that affects public trust in the scientific process~\cite{clark_university_2021, hallinan_unexpected_2020, jonathan_mayer_note_2021} and can erode members' trust in the community~\cite{hallinan_unexpected_2020}.

It is crucial for researchers to center ethical research practice for online communities, not only to avoid harms but to ensure that research has positive impacts on communities as well. 
In contrast to traditional offline communities, online communities present unique challenges related to the volume of studies, access to data and members, and geographically distributed members.
While Community-Based Participatory Research (CBPR) offers one model for ethical engagement with communities, its long-term commitment and emphasis on community-driven research questions can be a barrier for uptake~\cite{israel_critical_2017}.
Absent of a systematic alternative to CBPR, researchers have considered the effects of their work on longer-term community “health” considering factors such as composition, activity level, commitment, retention, and turnover of the community as a whole~\cite{ren_agent-based_2014, masli_evaluating_2012, yu_out_2017, 10.1145/1459352.1459356}. 
Much CSCW work bears on positive outcomes of research on online communities, through both empirical findings but also in potential for design implications to improve the quality of work. 
We find that these longer-term considerations and positive impacts may often be overlooked at the onset of research projects, and it remains unclear whether community members share the same understanding of community health, research impact, and what positive outcomes researchers could bring to these communities.

In this paper, we address this gap by proactively asking online communities about how they want research conducted about and on them. We conducted 9 community-engaged workshops with several critical online communities (Wikipedia, InTheRooms, CaringBridge, and r/AskHistorians) to develop a framework for ethical conduct of research with online communities.
We adopt a broad definition of community-level harm/benefit--an adverse/desired effect on the community--to allow for varying interpretations shaped by each community’s unique context (see Table~\ref{tab:participants}).
This paper centers the voices of members and organizers to answer two overarching research questions:
\begin{description}
        \item[\namedlabel{rq1}{RQ1}] How do online community stakeholders conceptualize \textit{community-level} harms \\and benefits?
        \item[\namedlabel{rq2}{RQ2}] How should decisions about online community research be made to address \\\textit{community-level} impacts?
\end{description}

\begin{figure*}[h]
\includegraphics[width=\textwidth]{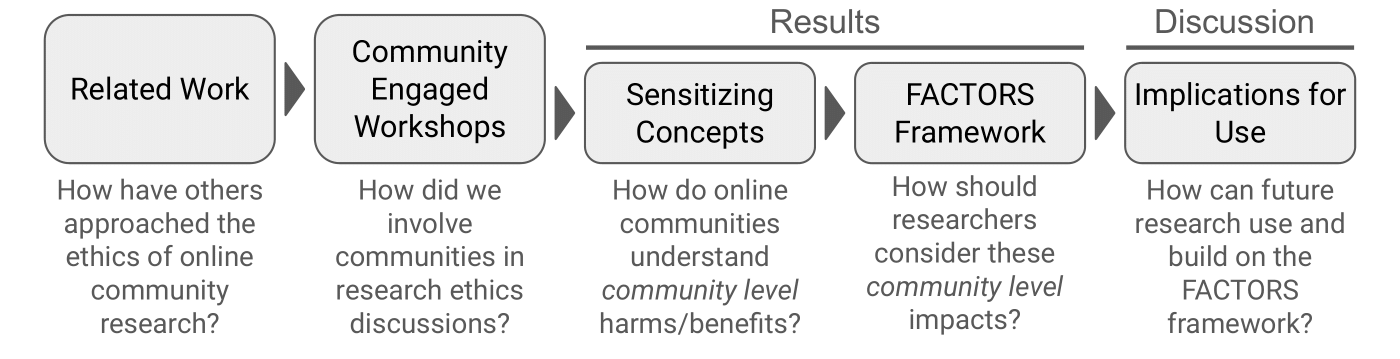}
\caption{
Paper outline: using community-engaged workshops to understand how research should avoid community-level harms and enhance community-level benefits.}
\label{fig:roadmap}
\Description{
Related Work: How have others approached the ethics of online community research? Community Engaged Workshops: How did we involve communities in research ethics discussions? Results: How do online communities understand community-level harms/benefits? FACTORS Framework: How should researchers consider these community-level impacts? Implications for Use: How can future research use and build on the FACTORS framework?
}
\end{figure*}

Figure \ref{fig:roadmap} outlines the paper and the primary questions we addressed. We found that community-level impacts of research directly connect to the communities' ability to achieve their goals; in some cases, research interferes with these goals, whereas in others it supports them.
Furthermore, communities frequently rely on individual or small groups of ``guardians'' to protect from community-level research harms.
Based on our findings and synthesizing our participants' perspectives, we propose the FACTORS (Functions for Action with Communities: Teaching, Overseeing, Reciprocating, and Sustaining) framework, a framework for online communities and researchers to avoid harms and enhance benefits.
These functions include overseeing subjective community rules, sustaining community resources, reciprocating research value to the community, and teaching researchers community norms and expectations.
The framework is designed to support ethical community research engagement and is most applicable to well-intentioned research that aligns with the values of the communities studied, rather than cases where community interests fundamentally conflict with research goals.
Building on a tradition that places a higher ethical burden on researchers, we argue that researchers should take a primary role in performing these functions to support their own and future online community research.

This research contributes the following to the CSCW and social computing research community:
\begin{itemize}
    \item We present an ethical framework (FACTORS) for researchers and communities to reinforce community-level benefits and avoid community-level harms. We have made the takeaways for this framework accessible through the FACTORS checklist.
    \item We provide empirical insights about online communities' understanding of community research.
    \item We provide guidance for researchers, online communities, and community organizers about how to conduct more ethical research in online communities. 
\end{itemize}

\section{Related Work}

In this section, we review prior academic and industry approaches to considering the ethics of research with online communities, articulating the key approaches and challenges to be addressed.

\label{sec:related_work}
\subsection{Ethical Research Principles and Policies}
While one can find statements about the need for research to benefit humanity dating back 500 years (e.g., to Francis Bacon’s Novum Organum~\cite{noauthor_novum_nodate}), specific treatment of rights of research subjects dates back to the Nuremberg Trials (and its finding of unethical conduct of German doctors). The Nuremberg Code of 1947 set down ten principles for permissible medical experiments, including voluntary informed consent, avoiding unnecessary suffering, and risk being no greater than the potential benefit~\cite{group_nuremberg_1996}. These principles were later codified and refined by the World Medical Association in the 1964 Declaration of Helsinki (revised periodically)~\cite{noauthor_wma_nodate} which remains in use as a legal basis for Good Clinical Practices. In the U.S., the Tuskegee Syphilis Study (1932-1972) and its withholding of penicillin treatment from unknowing African-American subjects combined with high-profile traumatic psychology experiments (including Zimbardo’s Stanford Prison Experiment~\cite{zimbardo_ethics_1973} and Milgram’s Obedience Studies~\cite{baumrind_thoughts_1964, milgram_behavioral_1963}), led to the passage of the National Research Act of 1974. This established Institutional Review Boards (IRBs) to review biomedical and behavioral research with human subjects and established a National Commission for the Protection of Human Subjects of Biomedical and Behavioral Research. The Commission's report, known as The Belmont Report~\cite{protections_ohrp_belmont_2010}, articulated three ethical principles: respect for persons, beneficence, and justice. These principles, and more detailed guidelines, form the basis of today’s Common Rule which governs research with human participants in the U.S.

These principles and policies have been quite effective at protecting individual research subjects, but they were never designed to address the impact of research on broader online communities—those who are not the direct subjects of an investigation, but who may be negatively affected by that research.
In the area of computer and information science, there are growing concerns that the Belmont Report and overview of ethics review boards are insufficient to manage the growing complexity of pervasive data research~\cite{shilton_excavating_2021}, online community research~\cite{hudson_go_2004, pater_no_2022}, and gaps in coverage with documents such as the Belmont Report~\cite{vitak_ethics_2017, vitak_beyond_2016}. A number of prior research threads have pointed to this substantial gap (e.g.,~\cite{kantanen_hazy_2016}) and it has been a recurring topic of discussion at workshops and community meetings on online ethics (i.e., the Aspen Institute~\cite{aspen_conference}, Future of Privacy Forum~\cite{future_bridging}, and Ada Lovelace Institute~\cite{ada_looking}). In this work, we applied community-engaged approaches to extend the consideration of risks and benefits of computational investigations to include community-level benefits and harms.

\subsection{Approaches to Community-Engaged Research}
Across disciplines, researchers have devised ways to more effectively and ethically engage with communities. One overarching practice across fields is community-based participatory research (CBPR)—common in public health research~\cite{mikesell_ethical_2013}. In CBPR, researchers are encouraged to join and participate in communities and leverage their skills to help solve community-identified needs and solutions~\cite{minkler_community-based_2008, strand_principles_2003}. These are long-standing partnerships that center the needs of the community and include community members in all aspects of the work, including generating research questions, implementation of the research study, and assessment. 
The approach of CBPR provides important insights for our study. For example, in \citeauthor{mikesell_ethical_2013}’s literature review, they outline components of ethical research engagement that move beyond the Belmont report, including key values such as reciprocity, mutuality, social action, and joint ownership~\cite{mikesell_ethical_2013}. 
However, CBPR’s emphasis on long-term collaborations may not be feasible in all research contexts (i.e., researcher-driven lines of inquiry).
In our work, we seek to identify alternative mechanisms for aligning community expectations with research practices to achieve similar ethical and participatory benefits.

In HCI, participatory approaches are a common source for involvement and engagement through design, such as participatory design (e.g.,~\cite{muller_curiosity_2014}) and value-sensitive design (e.g.,~\cite{friedman_value_2013}). 
Recent work has also highlighted how to make more equitable partnerships through community-based design opportunities~\cite{harrington_deconstructing_2019, le_dantec_strangers_2015, ross_human_2010}. 
However, research in ethical CBPR or participatory approaches in HCI often ask how to do participatory research more ethically – in other words, the development of these ethical practices may not cover non-participatory research and has typically focused on the perspectives and concerns foregrounded by the researchers themselves (e.g.,~\cite{fiesler_exploring_2016, vitak_beyond_2016}). 
In this work, we sought to apply the best practices and approaches of these participatory methods to develop a community-centered framework for research engagement with online communities.

\subsection{Rules for Engagement in Communities}
We are not the first to consider the rules of engagement with online communities. Several researcher-driven investigations have generated sets of rules and guidelines for online research. Two prominent ones include Communication Technology Research by The Menlo Report~\cite{noauthor_menlo_nodate} and Ethical Guidelines for Internet Research by AoIR~\cite{noauthor_internet_nodate}. 
While these provide general guidelines, they were shaped via dialogues between researchers, not through engagement with online communities, and may fail to account for the specific practices and considerations of those communities. 
For example, online recovery communities have a substantially different interpretation of the concept of “anonymity” than computer science researchers~\cite{rubya_interpretations_2017}, and these mismatches in understanding can lead to ethical breaches. 

The work on establishing community-driven guidelines has mostly occurred outside of the research context. Most sites already express some specific terms of service and sets of rules that may limit certain forms of research engagement. Such policies generally focus on limiting commercial exploitation on the site (i.e., spam) by prohibiting the use of automated tools for posting messages or automated tools for gathering data~\cite{fiesler_no_2020}, but it is also becoming more common for these pages to include information on acceptable research practices (e.g., X’s “Rules and Policies” page~\cite{noauthor_rules_nodate}). A few online communities have established explicit rules and practices for research engagements. For instance, to minimize privacy risks resulting from Tor-related research, Tor established the safety board which offers general safety guidelines and provides feedback for potential studies~\cite{noauthor_research_nodate}. Similarly, in 2020, Mozilla established Community Participation Guidelines~\cite{noauthor_community_nodate} which laid out a set of practices aimed at supporting the active participation of its members. In other cases, these rules come not from site creators, but from the specific moderators of a sub-community (e.g., r/EDAnonymous~\cite{chubmcgibbins_submitting_nodate}). 
Wikipedia is a pioneer in this space, establishing dedicated processes~\cite{noauthor_researchindex_2025}, how-tos~\cite{noauthor_wikipediaethically_2018}, and policies to guide ethical research engagements~\cite{noauthor_wikipedia_notlab_nodate}.
This makes Wikipedia a key collaborator in our work, as we seek to understand which practices effectively support communities and offer recommendations for communities with less research experience.

However, we observed limitations with existing online community rules. First, these communities often establish ad-hoc rules only as a reactive response to a violation breach where damages may have been already caused to the communities. 
Due to this reactive approach, many prominent communities do not have existing rules or processes in place. 
Second, existing rules may not be able to handle exceptions and the dynamic nature of ever-evolving research practices and methods  (leading some researchers to choose to disregard terms of service, viewing them as irrelevant or not in the interest of the community members~\cite{halavais_overcoming_2019}). 
Even Wikipedia's well-established research practices are not universally followed, suggesting a gap in the broader research community's uptake of community standards for ethical research.
We build on this understanding by working with both members and organizers of online communities to create a framework for the ethical conduct of research and implications for implementing this framework that could be proactively deployed by online communities.

\section{Methods}
\label{sec:methods}

To create a framework for ethical research with online communities, we conducted a series of workshops with four diverse online communities. For each community, we partnered with at least one organizer (e.g., moderator or employee) to advise and help coordinate community-specific norms for research engagement. We facilitated workshops inspired by participatory design with various community stakeholders. In this section, we describe our partnering communities, workshop recruitment, participants, workshop design, and analysis techniques.

\begin{table}
\caption{A table describing our participating online communities and who participated in each remote workshop. We leverage ~\citeauthor{10.1145/3637310}'s community archetypes to describe each community. The research dimension of difference describes the communities' existing practices for external research.}
\setlength\tabcolsep{5pt} 

\begin{tabularx}{\textwidth}{@{} p{0.15\textwidth} p{0.11\textwidth}@{}p{0.09\textwidth} p{0.4\textwidth} p{0.25\textwidth} @{}}
\toprule
\textbf{Community} & \multicolumn{2}{p{0.2\textwidth}}{\textbf{Description}} & \textbf{Dimensions of Difference} & \textbf{Participants}* \\
\midrule \addlinespace[.15cm]
Wikipedia & \multicolumn{2}{p{0.2\textwidth}}{Wikipedians edit, create, and discuss encyclopedia articles.} & 
\begin{tabular}[t]{@{}p{0.4\textwidth}@{}}
\textbf{Size}: 48 million registered editors\vspace{.15cm} \\
\textbf{Archetype(s)}: Content Generation\vspace{.15cm} \\
\textbf{Moderation}: Layered community moderation\vspace{.15cm} \\
\textbf{Research}: Publishes  information pages, an email list, and project proposal pages for researchers
\end{tabular}
& \begin{tabular}[t]{@{}p{0.25\textwidth}@{}}
Phase 1a** \\ 
n=7m \\ 
\rule{0.175\textwidth}{.25pt} \\
Phase 1b** \\ 
n=8m \\
\rule{0.175\textwidth}{.25pt} \\
Phase 2 \\ 
n=12 (8m, 6o, 5r)
\end{tabular} \\ \\
\multicolumn{2}{p{0.26\textwidth}}{Open Acknowledgments:} & \multicolumn{3}{p{0.65\textwidth}}{Olufemi Samuel, Matthewvetter, Zachary Levonian, Laura Couch, Nwoyeka Charles Chiemerie, and Lane Rasberry} \\
 \midrule \addlinespace[.15cm]
CaringBridge & \multicolumn{2}{p{0.2\textwidth}}{Patients and caregivers post health updates to personal journals usually shared with friends and family.} & 
\begin{tabular}[t]{@{}p{0.4\textwidth}@{}}
\textbf{Size:} 320,000 daily visitors \vspace{.15cm}\\
\textbf{Archetype(s):} Social Support, Learning \& Broadening Perspective, Content Generation \vspace{.15cm}\\
\textbf{Moderation:} Minimal site moderation \vspace{.15cm}\\
\textbf{Research:} Coordinated through non-profit employees
\end{tabular}
& \begin{tabular}[t]{@{}p{0.25\textwidth}@{}}
Phase 1 \\ 
n=7m \\ 
\rule{0.175\textwidth}{.25pt} \\
Phase 2 \\ 
n=7 (5m, 2o, 2r)
\end{tabular} \\ \\
\multicolumn{2}{p{0.26\textwidth}}{Open Acknowledgments:} & \multicolumn{3}{p{0.65\textwidth}}{Zach Levonian and Helen E Hansen}\\ 
 \midrule \addlinespace[.15cm]
InTheRooms & \multicolumn{2}{p{0.2\textwidth}}{People in recovery from substance use disorders exchange messages and attend virtual video 12-step meetings.} & 
\begin{tabular}[t]{@{}p{0.4\textwidth}@{}}
\textbf{Size:} 1 million members\vspace{.15cm} \\
\textbf{Archetype(s):} Social Support\vspace{.15cm} \\
\textbf{Moderation:} Moderated video meetings\vspace{.15cm} \\
\textbf{Research:} Monetized and coordinated through site employees
\end{tabular}
& \begin{tabular}[t]{@{}p{0.25\textwidth}@{}}
Phase 1 \\
n=5m \\ 
\rule{0.175\textwidth}{.25pt} \\
Phase 2 \\ 
n=9 (7m, 2o, 2r)
\end{tabular} \\ \\
\multicolumn{2}{p{0.26\textwidth}}{Open Acknowledgments:} & \multicolumn{3}{p{0.65\textwidth}} {Khadijah, Richard, and Rick\_011323} \\ 
 \midrule \addlinespace[.15cm]
r/AskHistorians & \multicolumn{2}{p{0.2\textwidth}}{Members ask questions about history and provide academic-level answers.} & 
\begin{tabular}[t]{@{}p{0.4\textwidth}@{}}
\textbf{Size:} 2.1 million members\vspace{.15cm} \\
\textbf{Archetype(s):} Topical Q\&A, Content Generation\vspace{.15cm} \\
\textbf{Moderation:} Actively moderated\vspace{.15cm} \\
\textbf{Research:} Coordinated through mods
\end{tabular}
& \begin{tabular}[t]{@{}p{0.25\textwidth}@{}}
Phase 1 \\ 
n=3m \\ 
\rule{0.175\textwidth}{.25pt} \\
Phase 2 \\ 
n=10 (2m, 3o, 6r)
\end{tabular} \\ \\
\multicolumn{2}{p{0.26\textwidth}}{Open Acknowledgments:} & \multicolumn{3}{p{0.65\textwidth}}{Maya Burak and bug-hunter} \\ 
\bottomrule
\begin{tabular}[t]{@{}p{\textwidth}@{}}
*: m, o, and r are members, organizers, and researchers; totals exceed sample size due to role overlap.
**: High interest from one Wikiproject led to a second workshop for broader generalization.
\end{tabular}
\end{tabularx}
\label{tab:participants}
\end{table}

\subsection{Stakeholders and Recruitment}
\label{subsec:recruitment}

The participating communities in this work were selected because they had prior experience with community research.
Each represented different context domains, archetypes~\cite{10.1145/3637310}, membership, and contribution structures (\autoref{tab:participants} highlights these differences).
Notably, each community had distinct approaches for navigating research which allowed us to examine a range of perspectives on ethical community-research interactions.
Our decision to structure this work around participatory workshops required high levels of member and organizer engagement, which constrained our sample to communities interested in collaboration (see Section~\ref{subsec:limitations}).
We collaborated with community partners—organizers in the community we've worked with in prior research (see Section~\ref{subsec:positionality})—to recruit for each workshop. 
We recruited three stakeholder groups from each community. Notably, these groups are not mutually exclusive. 
First, we define these broad groups:

\begin{enumerate}
  \item Community \textbf{Members} are people who participate in this online community. We define participation as engaging with community media at least monthly (i.e., reading posts, joining video meetings, or contributing content). Stakeholders in this group must self-identify as members and were recruited through community-specific mediums (i.e., mailing lists or recruitment posts).
  \item Community \textbf{Organizers} are people who build and/or maintain the community’s platforms, resources, and/or rules. Between online communities, organizers are very diverse and hold a variety of roles. In flattened community structures, we leverage the employees or volunteers who maintain the platform’s infrastructure. In hierarchical communities, we define organizers as moderators or leaders. Stakeholders in this group were recruited by invitation through our collaborating community partner.
  \item Community \textbf{Researchers} are people who actively conduct or have historically conducted research with the partnering community. Researchers are identified by their publication history or documented ongoing work with people or data from the community. Stakeholders in this group were recruited by invitation over email.
\end{enumerate}

We recruited 64 distinct participants across nine workshops\footnote{This total considers participants who attended multiple workshops as one distinct participant.} (described in detail in Table~\ref{tab:participants}). Recruitment with each community was split into two phases. During phase 1 workshops, we only recruited members to allow participants to reflect on broad community questions with others who may also have little to no experience with research in their community. This was done to acknowledge potential power differentials between stakeholders. Members who participated in phase 1 were re-invited as community research experts for their community's phase 2 workshop with members, organizers, and researchers. 

\subsection{Procedure} 
\label{subsec:design}

\subsubsection{Pre-Workshop Asynchronous Activities}

We used a pre-workshop survey to facilitate workshop scheduling with potential participants. The survey contained a consent form, candidate dates for the workshop to select from, demographic questions, a community involvement question, and two onboarding questions about community values and community research. These priming questions served two primary purposes: 1) allowing participants to asynchronously reflect on workshop topics, and 2) seeding phase 1 activities with examples grounded in the community. Then, participants were asked to reflect on the type of trust the community-research partnership had (see Appendix~\ref{app:sec:community} for more details)~\cite{wallerstein2020engage}.

\subsubsection{Workshops}

Each remote workshop was 2 hours and was composed of introductions, an icebreaker, three 20-minute breakout room activities, and a video of examples of community research\footnote{Each video contained 4 plausible examples inspired by past research in the community. They were scripted by two authors familiar with the community and performed by volunteers.}. Breakout rooms had the same 2-5 participants across activities and were balanced by stakeholders. Within each group, participants used Google Jamboard to produce a sticky-note-based artifact to respond to the activity's guiding question (See Supplemental File \textit{Community Research Workshop Materials}). In order, the breakout room sessions were: 1) \textit{What values are important to the broader \_\_\_ community?}, 2) \textit{What benefits/harms should future research strive for/avoid on \_\_\_?}, and 3) \textit{How should people make decisions about future research on \_\_\_?} In phase 1, the activities contained example sticky notes with common responses from participant's individual responses in the pre-workshop survey (i.e., 7 of 8 r/AskHistorians pre-workshop respondents talked about the quality of community answers, so one of the example values was accurate and rigorous information). In phase 2, we used salient themes from that community's phase 1 workshop. Otherwise, the content of phases 1 and 2 were identical.

The first author served as the primary workshop moderator and up to two co-authors acted as breakout room moderators for each workshop depending on the number of participants. Each workshop also had a technical moderator who helped participants with technical issues during the workshop. During breakout sessions, the moderators' primary job was to be a resource for participants (i.e. clarified questions and encouraged participation from all). All moderators shared their community positionality with participants, or lack thereof, during the introduction portion of each workshop. Supplemental file \textit{Workshop Moderator Discussion Guide} offers a step-by-step reference of our workshop protocol.
 
\subsubsection{Post-Workshops Asynchronous Activities}
\label{subsubsec:post}

After each workshop, participants submitted a post-workshop survey to receive their \$40 compensation and optionally evaluated their community's research relationship with the [Institution of Authors]. 
These evaluations were based on four dimensions of the Community Engagement Survey for evaluating CBPR processes (additional information and statistics available in Appendix~\ref{app:sec:community})~\cite{wallerstein2020engage}.
Immediately following the workshop we used member checking to validate our interpretations of participants' opinions~\cite{birt_member_2016}. The first author listened and wrote memos for each breakout session and generated a preliminary results document summarizing discussions to share with participants for comments and iteration. Inspired by CBPR where community partners play an integral role in helping adapt the research design to better suit the community~\cite{collins_community-based_2018}, we made small adaptations to our design at the recommendation of our community partners. For instance, on Wikipedia and r/AskHistorians, the preliminary results document was shared publicly via community channels for research for broader participation from interested community members who couldn't join the synchronous session.

\subsection{Qualitative Content Analysis}
\label{subsec:analytic}

\begin{figure*}[h]
\includegraphics[width=\textwidth]{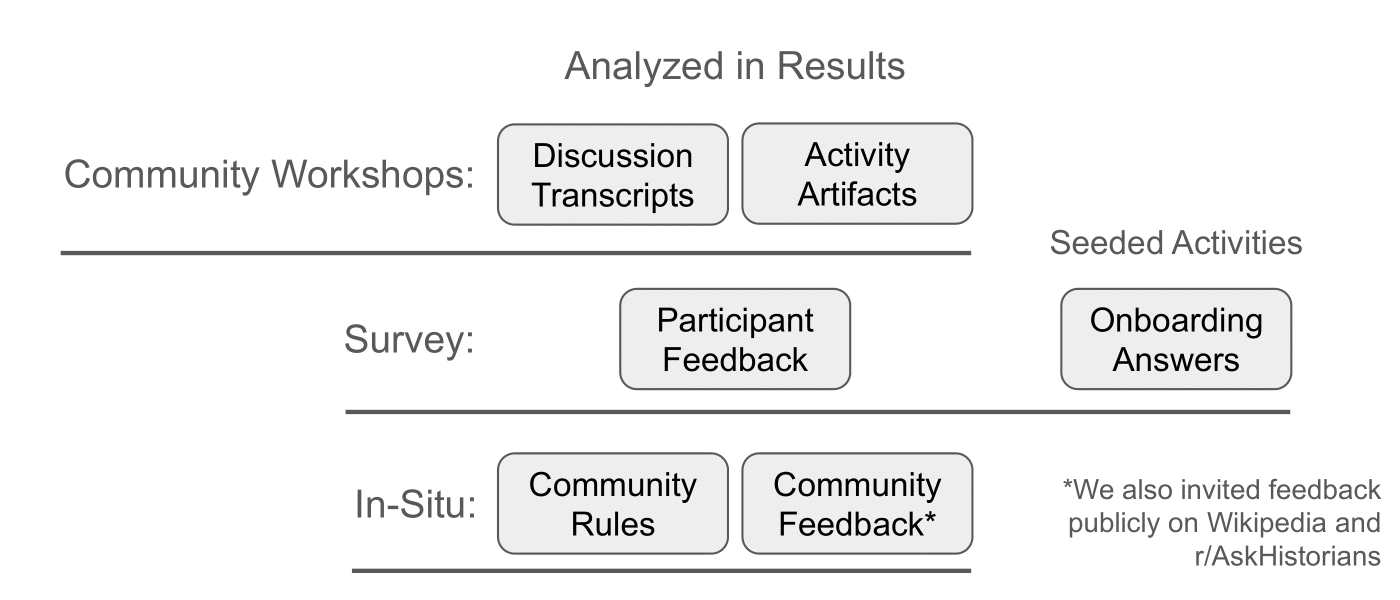
}
\caption{
Sources of qualitative data. Participants' onboarding survey answers were used to seed workshop activities, Discussion transcript, activity artifacts, participant feedback, community feedback, and community rules were analyzed to inform our results.}
\label{fig:data}
\Description{
Related Work: How have others approached the ethics of online community research? Community Engaged Workshops: How did we involve communities in research ethics discussions? Results: How do online communities understand community-level harms/benefits? FACTORS Framework: How should researchers consider these community-level impacts? Implications for Use: How can future research use and build on the FACTORS framework?
}
\end{figure*}

Figure~\ref{fig:data} details the six distinct sources we use to inform our community workshops and subsequent analysis.
Discussion transcripts, completed workshop activity artifacts, community rules documents, and participant results feedback~\footnote{We also invited feedback publicly on Wikipedia and r/AskHistorians at the suggestion of our community partner.}  were analyzed using data-driven thematic analysis to identify meaningful patterns across workshop activities and communities~\cite{muller_curiosity_2014}. This inductive process was chosen to understand the reality of community stakeholders affected by research and center their perspectives in our resulting framework.
The first author began by memoing and generating open codes from the transcripts, activity artifacts, and results feedback. Due to frequent references from participants to community rules for participation and research, these artifacts from the community were also analyzed in the context of workshop themes~\cite{noauthor_wikipediaethically_2018, noauthor_ah_rules_nodate, noauthor_aa_nodate}. This process generated over 400 open codes about community values, community-level harms and benefits, and stakeholder involvement in research decisions. The first and second authors worked together to discuss and cluster these codes by affinity until high-level themes emerged. These themes were presented to the entire author team to reflect their significance and novelty for the research community, including online community organizers.

\subsection{Ethical Considerations}
\label{subsec:positionality}

In this section, we consider how our identities, experiences, and values influence our work. The nature of this work and how it grew our partnerships with the participating communities in this study were the product of longstanding research relationships. It's important to acknowledge our positionality with the many communities that were involved in this work (see Table~\ref{tab:positionality}). We are mindful of how these past experiences helped us connect with participants, but also how they shaped what communities participated. All research activity was reviewed and approved by our university IRB, but IRB approval is not the end-all-be-all. Our team has first-hand experience with burned community relationships. We initially planned to include the Linux community but were advised against it by our community partner because of past community-level harms inflicted by our institution.

\begin{table}[]
\caption{A table describing authors' positionality in our participating communities before this work. Symbols r, m, and o denote a given author's identity as a community researcher, member, and organizer respectively.}
\begin{tabular}{llllllll}
              & \multicolumn{7}{c}{Author} \\
Community     & 1  & 2 & 3 & 4 & 5 & 6 & 7 \\ \hline
Wikipedia     & -  & - & - & r & - & r & - \\
CaringBridge  & r  & - & - & - & - & r & r \\
InTheRooms    & r & - & - & - & - & - & mr \\
r/AskHistorians & m  & - & - & m & - & - & - 
\end{tabular}
\label{tab:positionality}
\end{table}

One goal of this work is to center the perspectives of people in online communities at all phases of research. As such, we listened to community members' calls and gave participants the option to be identified by name or an identifiable pseudonym to credit them publicly for their contribution to this work in Table~\ref{tab:participants}. Participants were notified of this opportunity after receiving compensation along with resources outlining the permanence of this decision.
Prior work has set a precedent for deanonymization when the research involves intellectual contribution~\cite{bruckman_studying_2002}.

\section{Results}
\label{sec:results}
Our analysis of participants' reflections on online community research yielded ten themes spanning across communities.
Together, they reveal relevant aspects of research for communities' understanding of community-level harms and benefits (bolded about \ref{rq1}).
Broadly, these impacts relate to community trust, research equity, and capacity to achieve community goals.
We utilize these common features to form four sensitizing concepts~\cite{sanches2019hci} to orient the broader CSCW community to ethical online community research.
Participant quotes are attributed using the community abbreviation Wikipedia (Wiki), CaringBridge (CB), InTheRooms (ITR), and r/AskHistorians (AH) and stakeholder groups member (M), organizer (O), and Researcher (R).

\subsection{Research Needs to Align With Communities' Primary Purpose}
\label{subsec:values}

Aligning research with the community's primary purpose was a core element of participants' conceptualization of community-level harms or benefits.
The term \textit{primary purpose} is borrowed from the 5th tradition of Alcoholics Anonymous, ``The primary purpose of any 12-step group is to carry its message and give comfort to others who are still suffering''~\cite{noauthor_aa_nodate}.
This was discussed in the context of our InTheRooms workshops as welcoming newcomers, but similar motifs surfaced in other communities as goals or mission statements.
Participants had nuanced interpretations of their community's purpose, but broadly, it related to the reason they participate, and what distinguishes their community from similar others.
Stakeholders emphasized that research can either support or disrupt the community's purpose, depending on how well its methods, intent, and outcomes aligned with and actively contribute to the community's goals.

\subsubsection{Method Alignment}
\label{subsubsec:method_alignment}

When deciding whether a particular research project should or should not occur, participants assessed whether the project's methods would interfere with the processes and norms that sustain its purpose.
\textbf{Research methods preventing or distracting from members' ability to contribute to community goals are harmful to the community.}
Communities wanted passive (i.e. activity traces and data scraping) and active (i.e., interviews, interventions, and surveys) approaches to research that naturally integrated with community expectations and typical engagement patterns between members.

The way researchers collect data may have psychological effects on community dynamics.
A clear example emerged in the InTheRooms workshop.
Members pointed out that the acceptability of bots joining recovery meetings varied based on the type of information the bots collected.
Here, limiting the amount of data collected to specific words or behaviors maintained members' expectations that meetings are a safe and confidential space for sharing.
Notably, participating communities were unanimously opposed to methods involving deception (i.e.,  due to the risk of violating trust within the community.
This perspective illustrates a mismatch between the widely accepted use of deceit and incomplete disclosure in studies with minimal risk to individuals.

Several communities were concerned with active research methods \textit{``extracting engagement''} (Wiki/O/P1) from the community.
For instance, Wikipedia stakeholders did not want pop-up surveys or recruitment messages on community talk pages because they distract members and \textit{``hinder their process of actually editing'} (Wiki/M/P2).
Wikipedia's guide for researchers exemplifies this theme by emphasizing that ``methodology that interferes with the main goals of the encyclopedia is unlikely to get consent.''~\cite{noauthor_wikipediaethically_2018}.

\subsubsection{Intent Alignment}
\label{subsubsec:intent_alignment}

Participants wanted to understand not just how the research would be conducted, but who was conducting it, whether those researchers understood the community's goals, and how they intended to contribute to those goals. 
\textbf{Researchers that effectively communicate goals aligned to a community's purpose benefit the community by building trust in research.}
The closer the researchers' intended contribution matched a community's purpose, the more comfortable stakeholders were with the study. 

The researchers' intent was especially important when the perceived cost of research was high.
For example, in response to concerns about data scraping causing members to be less open about their health with the community, one CaringBridge member (P3) shifted their focus to the intent of the researchers:
\begin{displayquote}
    \textit{``Does the research follow the CaringBridge mission? ... What is the overall goal and if you're taking all of that public information, what is the goal of doing that?''}
\end{displayquote}
This perspective highlights a broader concern among communities that the information and resources extracted from the community meaningfully contribute to its mission.

One provocation about the difference between internal and external community research illustrates the importance of research's intended contribution.
The question of \textit{``Is [members designing tools for other members or internal research] in any way different from what we're talking about here with respect to [external] research?''} (Wiki/MR/P4) was raised across multiple community workshops.
For many, that distinction came down to the assumed goals of the researchers. 
When organizers conduct research in the community, they start with the premise of what \textit{``would benefit the most users''} (CB/O/P5). 
Members also believed organizers were more liable when things went wrong. 
Internal research was perceived to be more applied and translated to felt changes in the community. 

\subsubsection{Outcome Alignment}
\label{subsubsec:outcome_alignment}

Aligning research findings with community goals was crucial for demonstrating its value to the community.
Participants emphasized the importance of tangible benefits and expressed concerns about research that contradicted their goals.

\textbf{Research outcomes that actively contribute to the community benefit the community by building its capacity to achieve its goals.}
Members across communities were interested in research results that translated to felt benefits within the community
(i.e., increasing the quality of answers in r/AskHistorians or providing information to help new site authors get started on CaringBridge). 
When probed about the value of learning more about the community, participants were frequently interested in how research findings could improve outreach strategies and advocate for the community. 
As one InTheRoom member (P6) notes:
\begin{displayquote}    
    \textit{``I would say that's the benefit [of research for InTheRooms], because the more we know about how [InTheRooms] works, we may be able to get more newcomers to come in, once we know how to get them into InTheRooms.''}
\end{displayquote}
Similar research use cases can be seen on CaringBridge, where their landing page cites an external research survey—``CaringBridge users experience 3x greater feelings of connection to family and friends than the national average.''\footnote{2020-2021 ARCHANGELS National Caregiver Survey and Q4 2021 CaringBridge ARCHANGELS Survey of CaringBridge users.} 
Together, these positions echo the broader aim of increasing communities' capacity to reach new members.

\textbf{Research outcomes that misinterpret or are critical of a community's goals harm the community by driving away potential contributors.}
For instance, in the InTheRooms workshop, a new member proposed research that would screen members' desire to stop drinking before joining meetings.
Despite this being the only membership criterion for 12-step groups, other participants quickly highlighted how such a project misinterpreted the 3rd tradition\footnote{Tradition 3 of Alcoholics Anonymous states that the only requirement for A.A. membership is a desire to stop drinking} and would alienate newcomers, undermining the community's core values.
Similarly, on r/AskHistorians, we heard about research that took a critical stance on moderation in the community without considering the goals of the community. Here, misrepresenting the community in research can drive away new and existing members and provide a flawed scientific grounding for attacks on the community.

Sharing research findings that present a negative or critical stance on the community requires additional care to avoid community harms.
Wikipedia stakeholders from all groups emphasized that while uncovering issues within a community is a crucial first step, researchers must also consider how critical findings translate into tangible benefits for the community. 
They highlighted that exposing vulnerabilities or participation inequities without engaging with pathways for improvement risks exacerbating tensions rather than fostering change.
Similarly, InTheRooms participants discussed the importance of maintaining positivity in light of hypothetical research showing the community was not effective for sustained recovery to keep the community a viable resource for those in need in the interim.
Here, members across communities stressed the importance of research engaging with solutions to negative findings to avoid community harm, especially when the research is built on marginalized members' frustrations within the community.

Ultimately, aligning research methods, intent, and outcomes with the community's purpose and expectations is an important facet of stakeholders' conceptualization of community-level harms and benefits. 
We find that close alignment with these goals ensures research serves the community rather than disrupts it.

\subsection{Research Uses Community Resources}
\label{subsec:accumulating}

Community research often utilizes shared resources.
Individually, these small shifts may be incremental, but over time stakeholders described how they accumulate and impact community capacity to reach members, provide value, and reach its goals.
In practice, research can support community health on a larger scale by minimizing valuable resource use and providing tangible benefits in return to improve community perceptions of research.

\subsubsection{Resource scarcity}
\label{subsubsec:resource}

Participants cited various resources that were important to their community's sustained health.
While not exhaustive, discussions were centered around time, technical infrastructure, trust, attention, and effort.
Stakeholders were concerned with the immediate and gradual erosion of community resources that could ultimately reduce participation. 

Time was a particularly salient resource across communities, with participants citing internal motivators (i.e., volunteer work in a community is already time-consuming) and external motivators (i.e., community members are a vulnerable population and can't invest much time in the community) for its importance. 
Again, Wikipedia's researcher recommendations make this point explicit by saying, ``Studies should avoid interfering with the work of others, including by wasting volunteer time''~\cite{noauthor_wikipediaethically_2018}.
In one example, participants discussed a project that assumed they could survey 100,000 Wikipedia members only to realize part way through they couldn't reach their intended sample.
Here, misconceptions about the availability of community resources were doubly harmful because they used precious community resources and did not result in meaningful contributions to the community.

At scale, \textbf{the volume of studies using small amounts of community resources may become burdensome and constitute paper-cut harms.}
These frustrations compound with every new group of researchers, new study, and publication without materialized community benefit. 
Stakeholders were concerned with projects continuously \textit{``picking the same brains''} (Wiki/MO/P7) and differentially affecting community subgroups (i.e., admins or demographic minorities) by exhausting their time or effort.
\textbf{Research oversampling the same groups harms the community by creating echo chambers and risks core members being overloaded and feeling exploited.}
For the broader community, these ideas parallel the \textit{``tragedy of the commons''} (Wiki/O/P8) where the volume of unfiltered research access benefits researchers at the cost of the community.
To that extent, Lane Rasberry (P9), a Wikipedia member, researcher, and organizer, described an issue Wikipedia faces relating back to the important resource of community time:
\begin{displayquote}
    \textit{``Because there are more researchers wanting access to our community than we have community time to give the researchers. It's simply a scarce resource, and I think we should centrally manage access to the time of our community members in some fair way because it can't be a free-for-all where everyone grabs as much as they can.''}
\end{displayquote}

Research without valuable community contributions left stakeholders feeling like there was an opportunity cost to the research.
\textbf{Research that extracts community resources without providing value harms the community by diverting capacity from its goals.}
Building on this idea that the community's pride / time / effort could have built community capacity elsewhere one CaringBridge organizer (P10) said, \textit{``If resources are going to [research], what aren't they going to?''}
The prevalence of the perspective of research costs across participating communities illustrates the importance of demonstrating research value to the community.

\subsubsection{Research reciprocity}
\label{subsubsec:reciprocity}

Stakeholders were very positive about research minimizing and offsetting its costs by developing a culture of research reciprocity. 
While not a new concept, some participants felt there hasn't been a precedent for this for research in their communities. 
Awareness of tangible community research outcomes (see \ref{subsubsec:outcome_alignment}) was the primary way communities wanted research to give back.
Outside of research outcomes, organizers talked about ways for research to help balance its costs through donations to community-chosen causes to acknowledge the volunteer work required to support research. 
InTheRooms has taken this idea to the extreme by monetizing advertising studies to its members. 
Participants also described how different research methods, an upfront effort to understand the community, concise and well-written proposals, and better alignment of IRB evaluation with community values could minimize the cost of research on the community. 
This combination of minimizing cost and increasing the returns of research to the community builds a community's capacity to support research.

\subsection{Communities Seek Visible Accountability in Research}
\label{subsec:visible}

Visible accountability is the act of documenting and sharing information with the community at all stages of the research process.
Stakeholders wanted the researchers' intentions to be \textit{visible} to the community.
Documenting research activity for the community not only serves to hold researchers \textit{accountable} but also helps impact future research decisions.
Together, visible accountability can help prevent community-level harm and bring community-level benefits.

\subsubsection{Accessible Sharing}
\label{subsubsec:accessible}

Sharing research plans and outcomes for community members in accessible ways was an important aspect of visibility.
When the research was communicated clearly, participants felt better equipped to engage with the proposed research and findings.

At the proposal stage, participants wanted relevant aspects of the research concisely shared with interested stakeholders.
One AskHistorians organizer and researcher (P11) who frequently fields requests reflected on ways researchers could improve this process:
\begin{displayquote}
    \textit{``Make it clear in the ask how and why the research might be relevant to the community. Outline what kind of data is needed and plans for getting it. Summarize what is needed from moderators so they can evaluate if they have capacity.''}
\end{displayquote}
These points emphasize that researchers can minimize the labor involved in interacting with research and respect that this effort may be auxiliary to the community's goals.
Similarly, participants highlighted how using templates to document research simplifies the cognitive effort required to understand it by creating a shared language between stakeholders and projects.

Participants discussed a tension between the desire for increased visibility of research and for research to not interfere with other community goals (paralleling \ref{subsubsec:method_alignment}).
On one hand, stakeholders acknowledged that notifying all members of community research was not desirable. 
Still, in communities like Wikipedia with existing mechanisms for researchers to share proposals broadly (i.e. the meta-wiki research index), many felt they were not central enough to be noticed by members who might be interested.

How research findings are shared back to the community is equally important.
\textbf{Accessibly communicating research results benefits the community by enabling the findings to be understood and applied by the community.}
Conversely, pay-walled findings are unlikely to reach the community and bring benefits.
A step further, participants wanted research results to be written for the community.
Members saw value in working with community organizers to translate academic results into community results (see \ref{subsubsec:outcome_alignment}) to make them accessible in this way.

\subsubsection{Documenting for Future Research}
\label{subsubsec:document}

Stakeholders emphasized that documenting within the community has the added benefit of guiding future research efforts.
In this way, preserving the history of community research—good or bad—improves the quality of community research as a whole.

Participants discussed a reflective approach to documenting where the community could learn from research missteps.
They saw value in circling back on null results and unpopular research to generate new hypotheses or facilitate more scrutiny in that area going forward.
However, this was not a common practice for researchers in their communities.
In other ways, increased visibility in past research helps guide new studies to explore new questions instead of repeating the same perspectives.
\textbf{Established research findings and incremental gains are not valuable to communities and constitute harm by wasting resources.}
One novice Wikipedia researcher and newbie editor (P12) highlighted the significance of communal documentation of past research attempts and their outcomes:
\begin{displayquote}
    \textit{``Certainly, as somebody who might be a beginning researcher in a particular community, it's helpful to know there's a place to go where you can get some of those like, oh, this has been tried before and it's a bad idea, or this goes against our values, kind of feedback.''}
\end{displayquote}
Here, they underscored the value of visible feedback loops in the community research ecosystems to help steer new scholars away from approaches and research questions that may be harmful.

By accessibly documenting research and seeking input from those interested at all stages, researchers give stakeholders—members, organizers, and other researchers—the ability to shape and iterate on their work for the benefit of the community.

\subsection{Communities Rely on Guardians}
\label{subsec:guardians}

Guardians are community organizers who protect the community's goals by actively and passively mediating relationships between researchers and the community.
In practice, our participating communities had a small group or one interested organizer who interacted with community research to allow other stakeholders to focus on the internal community.
Members trusted guardians to make decisions about research on behalf of the community because they are uniquely positioned in the community to \textit{``see it all''} (CB/M/P13).
Additionally, members felt guardians may have more expertise to understand research risks when making these decisions.
Communities rely on these guardians as \textit{``pseudo IRBs''} (Wiki/MO/P7) because they find existing review mechanisms don't fully capture their interests.

\subsubsection{Active strategies}
\label{subsubsec:active}

Guardians actively prevent community-level harms by fielding research requests and collaborating with researchers.
Through this involvement, guardians can ensure that research aligns with the community's values and build community trust in research.

A subset of organizers discussed managing community research requests to protect community interests.
Guardians consistently wanted to stay informed about ongoing research, even when not required by ethical review.
Stakeholders highlighted a tension between the effort required to mediate community research and guardians' other roles.
In communities where this volunteer labor is at odds with other responsibilities, guardians' capacity to field research requests fills quickly.

Guardians can build the community's confidence in research by increasing research awareness and mutual understanding. 
Participants discussed how collaborating with guardians across different research phases can increase community trust by proxy of a trustworthy stakeholder.
Elaborating on this idea, one CaringBridge member (P14) envisioned actively onboarding researchers to help facilitate conversations and a shared understanding between researchers and community members: 
\begin{displayquote}
    \textit{``So could someone go and use my public site to do research without my consent? I mean, ultimately, they could. ... So that's kind of where I would see CaringBridge being vital in having some sort of onboarding for anyone who wants to do research around the Caring Bridge community. What are our values? Let's better understand the people whose lives you're entering into, so that you understand what they're going through, to an extent. ... Then CaringBridge could tell us as users of the site, research may occur, but we have stood firm that the researchers who are doing the research, understand our values and know what we're working towards.''}
\end{displayquote}
This example relates to other stakeholders' desires for guardians to act as stewards of community research by ensuring research alignment with community values and communicating its benefits to the broader community. 

\subsubsection{Passive strategies}
\label{subsubsec:passive}

Guardians also passively influenced community research by creating guidelines for researchers and clear rules for member engagement that also apply to researchers.
These structural safeguards reduce the need for direct involvement in research while ensuring that researchers are aware of the community's standards.

Wikipedia was the only participating community with documented guidelines for researchers. 
Organizers discussed previous efforts to create a formal research committee within Wikipedia responsible for screening research proposals.
This ultimately wasn’t sustainable given the volume and diversity of research requests the community received. 
Despite its dissolution, elements from the committee persist throughout the wiki as research guidelines.
This includes areas for researchers to pre-register their projects, recommendations for author positionality in the community, standards for appropriate (and inappropriate) methods, and statements about community values for open science initiatives.
While participants debated the pros and cons of reviving the research committee, they agreed that devoting effort to creating and maintaining community research rules was an efficient process to interface with research at scale.

Members across communities felt that community rules for participation had the ancillary benefit of protecting against active research methods that violated the community's core values.
For AskHistorians, members noted that researchers posting on the sub would be subject to the same moderator-enforced standards as members.
One Wikipedia member defined these as the community’s \textit{``bright-lines''} (Wiki/M/P15)—clearly defined rules—that apply equally to researchers and community members to ensure they respect the community’s norms.
Others stressed the community labor involved in moderating bring-line violations.
They emphasized that researchers understand a community's bright-lines before attempting to engage with the community.
\textbf{Research violating community rules harms the community by undermining its core values and creating work to reconcile breaches.}

\subsubsection{Democratizing guardianship}
\label{subsubsec:democratizing}

In some cases, guardianship wasn't limited to a small subset of organizers but instead democratized among members.
In this way, interested members can passively (i.e., by joint norm-setting) or actively (i.e., by providing feedback on proposed research) contribute to community research decisions. 

Member participation in maintaining community rules helps ensure the community's mission and self-proclaimed values represent its constituents.
Considering this, some participants explained how over time (i.e. \textit{``you're past the crisis moment''} - CB/MR/P16), a small percentage of members get more interested in the broader community (i.e. \textit{``how is this working or how can I make it better?''} - CB/MR/P16). 
In Wikipedia and r/AskHistorians, members described areas in the community where they could participate in these \textit{``meta-discussions''} (AH/M/P17).
In this way, community stakeholders can share the responsibility for sustaining a community's bright-lines.

Frequently, members envisioned being active consultants to help other guardians make more informed community research decisions.
Participants brainstormed a variety of processes to ensure members were the voice behind research decisions from voting to a representative \textit{``group conscious''} (AH/MO/P19).
In practice, Wikipedia encourages researchers to get consensus from the community during the pre-registering process for research proposals.
Stakeholders believed the more research engaged with members at this stage the less likely it would cause issues. 
These examples highlight how general rules for participation and guidelines for research—whether organizer or member-made—are important for guarding communities from community-level harms.

\section{Towards a Framework}

\begin{figure*}[h]
\includegraphics[width=\textwidth]{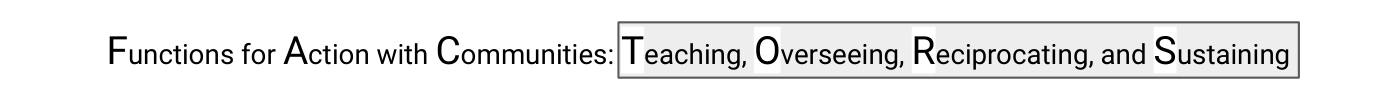}
\caption{
FACTORS Framework.}
\label{fig:factors}
\Description{
Functions for Action with Communities: Teaching, Overseeing, Reciprocating, and Sustaining
}
\end{figure*}

Leveraging the voices of several critical online communities, we guide our attention to how to ethically conduct online community research.
Mindful of the community-level impacts bolded in Section \ref{sec:results}, we synthesized the FACTORS framework (\autoref{fig:factors}) for ethical online community research (\ref{rq2}).

The framework identifies four functions designed to capture key features of the sensitizing concepts described in Section \ref{sec:results}.
We use the term \textit{function} to emphasize the broad, recurring roles that support ethical research engagement with online communities, rather than prescribing specific, rigid actions.
Each \textit{function} represents a key dimension of distributed cognition between researchers, online communities, and artifacts to navigate community-level research harms and benefits: 1) Overseeing community bright lines, 2) Sustaining community resources, 3) Reciprocating value, and 4) Teaching researchers community norms.

In the following sections, we first define each function in the context of our results.
Then, we provide examples of \textit{mechanisms} to be enacted by researchers and communities to provide concrete examples of FACTORS in practice (see Table~\ref{tab:synthesis_into} for an overview).
A \textit{mechanism} is an instantiation or implementation of a process that facilitates one or many functions for action with communities. 
By emphasizing artifact-based mechanisms we aim to reduce the engagement-heavy nature of community research ethics while still promoting ethical alignment and community agency.
Recall that our motivation for working with these four communities was to understand what practices work for them in order to make recommendations for others.
While we did not initially set out to identify gaps in existing norms, each community workshop revealed opportunities to enhance the visibility and influence of existing mechanisms while also reducing their burden on the community.
Grounded in these reflections, we discuss the effectiveness of example mechanisms on three dimensions: 1) visibility to researchers, 2) influence on research decisions, and 3) community involvement.

\begin{table}[]
\caption{Overview of the FACTORS framework (see \autoref{fig:factors}) illustrating example community-level impacts and mechanisms designed to address them. Mechanisms that perform each function should be identified in online communities before starting research to avoid community-level harms and reinforce benefits. Examples of harms and benefits were raised during community workshops.}

\label{tab:synthesis_into}
\centering
\rowcolors{1}{offwhite}{}
\begin{tabular}{>{\raggedright} p{0.15\textwidth} >{\raggedright} p{0.34\textwidth} >{\raggedright\arraybackslash} p{0.436\textwidth} }
\toprule
\rowcolor{white}
\multirow{2}{*}{\textbf{\centering Functions}} & \multicolumn{2}{c}{\textbf{Example Scenarios}} \\ \cmidrule(lr){2-3}
\rowcolor{white}
& \textbf{Without Mechanism} & \textbf{With Mechanism} \\ \midrule
Teaching Researchers Community Norms (\ref{subsubsec:teaching}) & On InTheRooms, researchers build a system to screen member's desire to stop drinking, misinterpreting the 3rd tradition. & If InTheRooms had an onboarding course for researchers to learn how members interpret each tradition, this could have been avoided. \\
Overseeing Bright-lines (\ref{subsubsec:overseeing}) & On AskHistorians, researchers promote new members unfamiliar with subreddit rules to post answers, leading to many mod-removed comments. & If an AskHistorians mod had reviewed the study, it could have been prevented.  \\
Reciprocating Value (\ref{subsubsec:reciprocating}) & On CaringBridge, authors find out their data was used in a study but don't know why, leaving them with less trust in the community. & If CaringBridge had a newsletter to share how the research impacted the community, it could have demonstrated its value. \\
Sustaining Community Resources (\ref{subsubsec:sustaining}) & On Wikipedia, researchers repeatedly sample the same minority group, causing them to burnout from the community. & If researchers had documented prior studies within the community, it would have been easier for future researchers to identify and redesign similar studies. \\
\bottomrule
\end{tabular}
\end{table}

\subsection{Teaching Researchers Community Norms}
\label{subsubsec:teaching}

Teaching researchers about community norms and values functions to offload the burden of making research decisions to researchers and hold research accountable for its actions.
Instead of making one-size-fits-all pronouncements, principles for ethical decision-making in internet research call for an inductive approach to understanding specific contexts~\cite{noauthor_internet2_nodate}.
Socializing researchers to unique community norms preserves the values that drive community participation over similar alternatives.

In principle, this function is deeply related to the others because it increases researchers' ability to make informed decisions that align with individual community expectations.
This simultaneously captures researchers investing more effort to proactively identify community-level impacts, and making the elements to inform those decisions more visible to researchers.
Communicating this shared understanding of community values in the intent behind research assures stakeholders that their interests are aligned.

\subsubsection{Mechanisms}
Communities can effectively educate researchers by encouraging active participation, documenting standard operating procedures (SOPs), and creating onboarding policies.
In practice, we find many communities rely on functions for socializing newcomers to also sensitize researchers to community norms.
Mission statements, getting started guides, and FAQs are document-based mechanisms to educate and have high visibility for all because they are central to the community's purpose. 
Communities also discussed researchers participating as active members to learn norms through trial and error and feedback from community members.

Alternatively, SOPs for research encapsulate the idea of artifact-based mechanisms for overseeing bright-lines (\ref{subsubsec:overseeing}), reciprocating value (\ref{subsubsec:reciprocating}), and sustaining resources (\ref{subsubsec:sustaining}).
As mentioned in prior subsections, these explicit guidelines can draw attention to important norms and values for research to adhere to.
Participants built on this idea by offering formal, organizer-driven onboarding courses for researchers, which teach them about community values and provide resources for conducting studies. 
While this mechanism requires more upfront involvement with communities to create, it has a strong propensity for influence and researcher visibility.

Educating researchers on community norms and communicating a shared understanding functions to protect community bright-lines, preserve important resources, and promote reciprocity in community research.

\subsection{Overseeing Bright-Lines}
\label{subsubsec:overseeing}

To avoid ad-hoc responses to harmful research, communities need mechanisms for guarding important community boundaries that should not be crossed. 
As the name suggests, these should be unambiguous policies for research that are central to the community's goals. 

When defining policies, prioritize preserving key aspects of the community that are difficult to restore (i.e., volunteer effort, trust, and stability). 
We find communities' bright-line rules often relate to how members, and by extension researchers and research methods, engage with the community. 

This function should clearly outline the repercussions of crossing community bright lines (i.e., loss of privileges or institutional involvement) and include resources for members to escalate non-compliant research to increase the function's influence. 
Here, member-driven moderation efforts in related CSCW research~\cite{10.1145/3415178} have a higher propensity to reflect community consensus.

\subsubsection{Mechanisms}
Communities have active (i.e., approval processes) and passive (i.e., rules for engagement) strategies for overseeing their bright lines. In communities with organizer-enforced rules like AskHistorians, a distributed process between organizers and guidelines may be ideal for overseeing community bright lines.
Their subreddit rules are visible to all visitors and encompass dos and don'ts for engagement (i.e. bot accounts and polls), and clearly defined outlets for connecting with moderators. 

For other communities, platform owners may choose to gate access to APIs to avoid bright-line violations.
While these gates serve to oversee bright lines, we note the growing voice in HCI research to work around platform terms of service~\cite{halavais_overcoming_2019, noauthor_internet2_nodate} and instead highlight the importance of being explicit about why observing bright lines is important to preserving community values~\cite{fiesler_no_2020}. 

These two examples present mechanisms dependent on community organizers' involvement, but that may not be desirable at scale or in communities with distributed oversight. 
Instead on Wikipedia, we see elements of overseeing in pre-registering research on the meta-wiki research index with member empowerment to oversee community bright-lines and policies for research~\cite{noauthor_wikipedia_notlab_nodate}.
These mechanisms' proximity to helpful resources for researchers increases their visibility, and clearly defined repercussions strengthen their influence. 

Overseeing community bright-lines can prevent research that interferes with online communities' goals and primary purpose.
Active strategies that engage individually with projects give communities more control over gatekeeping community bright-lines.
However, passive strategies may be more desirable at scale and emphasize the importance of clear and well-defined rules for research.

\subsection{Reciprocating Value}
\label{subsubsec:reciprocating}

Reciprocating the value of research to online communities simultaneously functions to foster trust in research, build community capacity, and equalize the beneficiaries of research.
This function supports sharing research outcomes back to the community, which by definition involves the community, but the onus is on the research to demonstrate its value to online communities~\cite{proferes_studying_2021}. 
Disparate positive impacts of community-based research have been an unresolved issue since the mid 1970s~\cite{green_unresolved} and continues to be a point of discussion in CSCW research today~\cite{cooper_systematic}.
For individuals, prior work has found that the value of community projects is determined by their ``day-to-day'' applicability for members~\cite{penuel_principles}.
For communities, our findings illustrate that using communities' goals to define and measure research outcomes can help bridge the gap between metrics relevant to online communities and researchers. 

We know researchers believe transparency in the intent and practice of research is an important quality of online data ethics~\cite{vitak_beyond_2016}, but this is challenging to accomplish without detracting from a community's primary purpose or without risk to researchers ~\cite{10.1145/3406865.3418589}. 
Importantly, functions for sharing research value should be performed with the community as the intended audience (i.e., accessible and actionable) to enable transparency while minimizing risks to researchers and the community. 

\subsubsection{Mechanisms}
This function involves creating mechanisms for research benefits to be shared and felt within the community.
In practice, creating appropriate spaces for meta-discussions simultaneously gives interested members and researchers a place to discuss \textit{``How can we do this better?''} (CB/MR/P16). 
Meta-community channels for engagement naturally lend themselves to research results sharing without interfering with a community's primary purpose.

Research newsletters are well-equipped to share results with an interested subset of members. 
Again, results should be concise and accessible to a broad community audience to enhance understanding.
Community research venues and recognition at those venues can serve as incentives for research with community-level impacts to increase the influence of this function.

Community stakeholders expressed concerns about the potential harms of sharing negative research results, fearing that such findings could harm the community's reputation and well-being, echoing the risks articulated in prior work~\cite{ehsan_negotiating_2024}. 
Drawing on best practices in related fields, the Cybersecurity and Infrastructure Security Agency encourages companies to establish policies for how researchers should notify and how long they should wait to publish vulnerabilities~\cite{cisa_vulnerability}.
Similar community policies or statements could serve to reciprocate the value of critical research while navigating potential negative impacts on online communities.
Such a mechanism exemplifies a path forward in situations when community and researcher incentives may be at odds.

Related to reciprocating value through organizer involvement is the member-perspective that organizers can act as stewards of research.
Recent CSCW work has proposed the active role of community ``mediators''—someone who vets the researchers—to communicate the value of community-research partnerships by proxy of trusted members~\cite{kotturi_peerdea}.
Alternatively, communities can create general resources for members about who researchers are and their value (i.e.,~\cite{noauthor_wikipedia_dontbite_nodate}) to normalize ethical community research.

Ultimately, demonstrating and communicating research benefits for the community builds community capacity to achieve its goals and offsets the paper cuts of research.
Effectively reciprocating benefits to members and organizers is important for ethically engaging with online communities.

\subsection{Sustaining Community Resources}
\label{subsubsec:sustaining}

Sustaining community resources across the ecosystem of research within a community is important for mitigating the compounding costs of research. 
This function serves to shed light on the amount of research the community has and is currently supporting to inform decisions about its capacity to field research. 

When the volume of research puts undue stress on a particular community resource or subgroup, each additional study amplifies the harm to the community.
\citeauthor{to_flourishing} found an over-focus across research studies on damaged-centered design for ethnic minority groups in online communities presented an incomplete narrative that was ultimately harmful for researchers and communities.
As such, mechanisms should interrogate the type, amount, and duration of resources needed to support a given project in the context of the broader body of research currently supported. 

Beyond current use, sustaining resources involves a critical reflection of past community research to understand what questions and subgroups in the community have been the target of prior investigations to avoid research burnout.
Moreover, research has argued that openly considering the history and context of research with communities builds trust and equity research engagements~\cite{harrington_deconstructing_2019}.

\subsubsection{Mechanisms}
Community artifacts that make proposed, current, and completed research visible can serve to sustain resources.
Pre-registering research proposals is likely the most complete example of an artifact-based process for this function in our participating online communities. 
Wikipedia's meta-wiki research index serves as a searchable living history of completed and in-progress research alongside useful resources for researchers like mailing lists, datasets, and conferences (i.e., increased visibility). 
Research pages are prepopulated with templates created by organizers to concisely highlight resources relevant to the community like research timelines, project status, funding, recruitment/data collection, and sample sizes. 
This practice creates a shared language between projects that simplifies evaluating proposals and increases the influence of this function.

Despite these strengths, the existing process can be improved. 
Stakeholders described how the research index is not universally utilized which suggests the need for it to be more central to how researchers access the community. 
We also see the opportunity for added visibility in failed and abandoned projects and recommendations for researchers to inform decisions concerning the ecology of community research. 

Community newsletters can facilitate sustaining resources by highlighting completed or ongoing community research.
Ideally, newsletters should be preserved to record the history of research.
This process may have a delayed influence on research depending on the frequency and process implementation.

The function of sustaining community resources across projects helps inform community research decisions that minimize the cost of research on the community.
By increasing visibility into current, past, and proposed research, studies can minimize the negative impacts of research on the community

\section{Discussion}
In the sections below, we reflect on our findings and FACTORS framework for online community research.
Alongside these four functions, we present implications for researchers and online communities to guide future work and designs that consider community-level impacts of research.

\subsection{Implications for Ethics Researchers}

First, we concentrate on implications for ongoing HCI community discussions around research ethics.
We extend this important body of work by involving a critical set of online communities and their stakeholders in broader community ethical discussions.
Specifically, we build on the HCI community's empirical understanding of the potential harm or risk for online communities in research~\cite{noauthor_internet_nodate} and contribute the FACTORS framework for navigating community-level impacts of research.  

In many ways, we find a strong overlap between communities' conceptualization of community-level benefits and harms and researcher-driven guidelines for internet research (i.e., data minimization~\cite{noauthor_internet_nodate}, action research~\cite{khanlou_participatory_2005}, and developing codes of conduct~\cite{zook_ten}).
Communities' nuanced interpretations of important ethical considerations support prior calls for a context-specific approach to considering the ethics of internet-based research ~\cite{noauthor_internet_nodate, fiesler_remember, dym_ethical, zimmer_addressing}.
We find that empowering communities to enact FACTORS (i.e., rules for research) may facilitate a deeper alignment to community norms and expectations than ad-hoc determinations about what is appropriate.
Importantly, mechanisms for enacting these functions are flexible for cases when contacting organizers may not be appropriate or possible.
Many existing online communities do not have the capacity to undertake or organize the aforementioned functions. 
We see the potential for researchers to find creative solutions that work for the communities they study.
In light of our results, future research should prioritize making it easier for online communities to identify their own guidelines for research and increasing the visibility of those unique considerations within the community. 

We find communities care deeply about researchers' responsibility to share benefits back to the community~\cite{noauthor_internet_nodate, proferes_studying_2021}.
Related to beneficence, sharing research results has been a prominent focus in online research ethics~\cite{gilbert_big, fiesler_participant_2018}.
Our findings indicate that accessible and actionable result-sharing is the primary way communities want research to reciprocate its value, which directly impacts community trust in research broadly. 
Trust in research, researchers, and research goals is important for realizing community-level benefits~\cite{mcdavitt_dissemination_2016, jagosh_realist_2015} and mitigating perceptions of harms~\cite{hallinan_unexpected_2020}.
Prior work has noted the risks to researchers and communities in sharing some research findings~\cite{fiesler_remember, massanari_rethinking, suomela_applying}.
Our findings illustrate the importance of maintaining positivity in communities when sharing research, particularly by engaging with solutions for findings critical of community goals.
However, our decision to prioritize community voices limited our ability to holistically consider risks to researchers.
This tension between the priorities of communities and researchers highlights the need for future work to understand effective strategies for reciprocating value while preserving the safety of researchers.

\subsection{Implications for Researchers of Online Communities}
Next, we focus on implications for researchers who do work involving online communities--which is a substantial portion of the CSCW and HCI communities. 

Online communities have a variety of goals, many of which do not involve supporting research. 
Participants talked about their communities existing to connect patients with families, maintain an encyclopedia, find support in one's recovery, and preserve public history.
As such, much of the responsibility is on researchers to care for FACTORS in online communities early and often in the research process.

When proposing projects, researchers should plan for contingencies that reduce the pressure to continue if they encounter friction from online communities.
As one organizer put it, \textit{``slow down and consider the challenges that may be ahead''} (Wiki/O/P1).
Plan for alternative communities in grants and proposals to reduce the demand to push through friction, uncertainty, withdrawn consent, or instability within a community.
Include tangible benefits for communities in the broader impacts section of grants and institutional reviews to better align academic incentives.
Another organizer shared, \textit{``even if you do everything right, sometimes the research still won't work (i.e., too much going on behind the scenes, not be enough incentive to make it happen, at capacity with other researchers, not enough active participants, etc.)''} (AH/MR/P11).
These points illustrate the importance of flexibility in research plans to respect the needs of online communities.

When proposing online community research, we have the responsibility to identify FACTORS in communities and, when appropriate, work with organizers to implement lightweight solutions. 
First, we offer a checklist for researchers entering into communities to evaluate in Appendix~\ref{app:checklist}.
This sample set of questions is intended to prompt reflection on the community-level impacts of the proposed research.
When mechanisms for supporting these functions aren't visible, researchers can contribute by establishing researcher-driven FACTORS (i.e., research mailing lists and meta-community wikis). 
Finally, we encourage researchers entering into new online communities to talk to stakeholders about community health and community-level impacts of research.
We provide replicable workshop protocols in supplemental file \textit{Community Research Workshop Materials} and \textit{Workshop Moderator Discussion Guide} to facilitate these discussions.

\subsection{Implications for Community Organizers}
In this section, we offer an executive summary of our results for community organizers.
Future research is needed to make this process easier for online communities. 
These recommendations serve to start dialogues between researchers and communities about community-level benefits and harms.
\begin{enumerate}
    \item Prioritize guides, \textit{bright-line} rules, and documentation for others to make research decisions that align with community research expectations. What level of community involvement is desired for different types of research (i.e., observational, intervention, interviews, big data)? What values are important to preserve and how do they contribute to the community's goals? 
    \item Members are curious about community values and value alignment between community stakeholders. Create spaces for stakeholders to have meta-discussions about community health. Many participants appreciated synchronous dialogues about these topics. We offer PDFs of our workshop materials and guides to facilitate workshop discussions in supplemental file \textit{Community Research Workshop Materials} and \textit{Workshop Moderator Discussion Guide}. What does a ``healthy'' community look like?
    \item Articulate the value of research for the community. Consider how external resources could be mutually beneficial. What measures of success are relevant to community goals?
    \item Educate community members on the nature and risks of contributing content they may not want to be used for other purposes. Communicate what community policies for public data use do and don't protect. What could enable members to have more control over their data sharing? 
\end{enumerate}

\subsection{Limitations and Future Work}
\label{subsec:limitations}

Our goal was to center the voices of several critical online communities in an ethical framework for ethical conduct with online communities using community-engaged participatory workshops. 
This methodological choice led to a variety of trade-offs and limitations, which we outline here to contextualize our findings and motivate future research.

We acknowledge the theoretical nature of the FACTORS framework and see the need for future research to develop strategies for evaluating the effectiveness of specific FACTORS mechanisms and assessing its real-world applicability. 
As a first step, this work presents three dimensions relevant to community stakeholders for evaluating FACTORS mechanisms--visibility, influence, and community involvement. 
Future work should continue to refine these evaluation dimensions and identify quantifiable metrics to facilitate robust development of FACTORS mechanisms.
Related to the FACTORS framework's real-world applicability, we identify a relevant gap in CSCW broadly for evaluating frameworks and theories in practice. 
~\citeauthor{halverson_activity_2002} presents four types of theoretical “power” relevant for future work to assess the efficacy of the FACTORS framework: descriptive (i.e., its ability to describe and make sense of the world), rhetorical (i.e., its ability to help in the communication by naming and categorizing concepts), inferential (i.e., does it help make inferences such as whether a particular approach is likely to lead to ethical problems), and application power (i.e., its ability to apply to the world and inform pragmatic decisions~\cite{halverson_activity_2002}.
To address application power, we introduce the FACTORS Checklist (see Appendix~\ref{app:checklist}) as an initial tool for researchers to apply the FACTORS in practice, but further empirical work is needed to assess its effectiveness in guiding ethical community research.

While our participating communities were diverse in terms of archetypes and each represented different approaches to and experiences with online community research, our decision to seek out communities with existing research relationships limited who participated in this work. 
This enabled our work to build on communities' prior experiences with research and break out of the status quo of ad-hoc rules in response to community breaches of trust.
At the same time, this approach means that our research and the FACTORS framework may not capture all contexts where community research occurs.
Particularly those with no prior experience with research, emerging communities, deviant~\cite{coletto2016behaviour} or illicit communities~\cite{10.1145/1964897.1964917}, politically polarized spaces, and decentralized networks--each with unique ethical challenges for community research.
Through the nature of our research, we only captured the perspectives of those willing to participate.
Accordingly, we recognize that our results do not represent the full demographics of the communities we studied.
In particular, we may not have captured the perspectives of community members who hold such negative views about research that they chose not to participate.

These limitations constrain the generalizability of our work but also motivate future work to further inform ethical research with online communities.
For example, because we have not engaged with communities organized around illicit or unethical activities, we do not yet feel comfortable recommending changes in the Common Rule or other legal frameworks governing human subjects ethics review.
Participating stakeholders highlighted the opportunity to better align IRB human-subjects review and determination with the expectations of online communities to prompt researchers to consider these factors explicitly.
The FACTORS framework is not intended to replace, or be the sole lens for evaluating, the ethics of community-research interactions.
Instead, we feel that review can be informed by FACTORS to better recognize respect for communities as well as for individuals. 
While FACTORS emphasizes alignment with community values, we acknowledge that not all online communities are receptive to research or engage with it in good faith.
As such, universal rulemaking would need to factor in whether and how to respect these communities.
Critically, there is a need for future research to collaborate with such communities to inform these changes.
To support replication in these contexts, we include 2 supplemental files: 1) \textit{Workshop Moderator Discussion Guide}, a detailed script for conducting a 2-hour community-research workshop, and 2) \textit{Community Research Workshop Materials}, slides for a collaborative whiteboard-style remote workshop~\cite{wilson_replichi_2011}.
We highlight the risks to researchers and communities when conducting research with communities whose values significantly differ from researchers.
To mitigate these risks, \citeauthor{10.1145/3555182} suggest designing for shared values to navigate the tension between community-level harms and researcher safety.
Future work can offer valuable insights by including new populations with diverse community experiences, particularly those with negative or skeptical views toward research while being mindful of potential community-level harms in more resistant environments.

\section{Conclusion}
Online communities are a central focus of social computing research, but high-profile breaches of community trust suggest a disconnect between existing ethical practices and community expectations for research.
We conducted nine participatory-inspired workshops with four critical online communities (Wikipedia, InTheRooms, CaringBridge, and r/AskHistorians).
We found four key aspects concerning how communities navigate community-level impacts of research: 1) research needs to align with communities' primary purpose, 2) research uses community resources, 3) communities seek visible accountability in research, and 4) communities rely on guardians.
Considering these facets, we leverage voices from these communities to synthesize an ethical framework (Functions for Action with Communities: Teaching, Overseeing, Reciprocating, and Sustaining) for avoiding community-level harms and reinforcing community-level benefits.
We conclude with implications for researchers to be the agents of change in identifying and implementing mechanisms to protect the online communities they study.

\begin{acks}
We offer a special thanks to our participating communities and community partners for their interest in this project. We would also like to thank Ruixuan Sun, Juan F. Maestre, Daniel Runningen, and Nuredin Ali for their contributions to our community research workshops. This work was supported by NSF ER2 Award \#2220509.
\end{acks}
\bibliographystyle{ACM-Reference-Format}
\bibliography{bibliography}

\appendix
\section{FACTORS Checklist}
\label{app:checklist}

Similar to prior work~\cite{noauthor_internet2_nodate, kaptelinin_methods}, Table~\ref{tab:checklist} details a sample set of questions for researchers to consider when entering into communities. 
These questions are intended to be a starting point for researchers to use to identify FACTORS in online communities.
Importantly, all the following provocations have the added element of \textit{and how do you know?} 
When the answer is unclear, consider collaborating with community guardians to make these mechanisms more visible for future researchers.

\begin{table}[]
\centering
\renewcommand{\arraystretch}{1.5}
\caption{Sample questions to help researchers identify functions for action with communities.}
\label{tab:checklist}
\begin{tabular}{>{\raggedright\arraybackslash}p{0.23\textwidth} | >{\raggedright\arraybackslash}p{0.23\textwidth} | >{\raggedright\arraybackslash}p{0.23\textwidth} | >{\raggedright\arraybackslash}p{0.23\textwidth}}
\multicolumn{1}{c}{\large\textbf{Teaching}} & \multicolumn{1}{c}{\large\textbf{Overseeing}} & \multicolumn{1}{c}{\large\textbf{Reciprocating}} & \multicolumn{1}{c}{\large\textbf{Sustaining}} \\
\toprule
\rowcolor{mygreen} 
\textbf{What are the community's norms for research?} &
\textbf{What bright-lines does this community have?} & 
\textbf{How will this research demonstrate its value to the community?} & 
\textbf{Does this community have the resources to support this work?}  \\
\midrule
What are the community's core values?

\vspace{.25cm}
What types of community engagement and research methods are appropriate in this context?

\vspace{.25cm}
What are the community's expectations for privacy irrespective of the publicity of community content~\cite{fiesler_participant_2018}?

\vspace{.25cm}
Who in the community has provided feedback on the proposed work?

\vspace{.25cm}
What is the research team's positionality within the community? &

What rules for research does this community have? And what rules do similar contexts have that may also apply?

\vspace{.25cm}
How does the community oversee its bright lines?

\vspace{.25cm}
Who has the ability and knowledge to share community concerns about this research?

\vspace{.25cm}
Who on the research team is accountable for bright-line violations?

\vspace{.25cm}
How will this research affect trust within the community~\cite{hallinan_unexpected_2020}? Trust in research~\cite{FEITELSON2023111774}?

\vspace{.25cm}
What contingencies are in place if this work encounters community bright-lines post-hoc? & 

How does this research benefit the community?

\vspace{.25cm}   
What aspects of this work are most relevant to the community? How will they be communicated to community stakeholders?

\vspace{.25cm}    
Where is it appropriate to share our research results intended for the community? Are they true? Necessary? Kind? \textit{Useful?}

\vspace{.25cm}   
Who in the community has the ability and interest to apply the intended contribution to the community?

\vspace{.25cm}    
How can the research team increase the visibility of this work without impeding on the community's primary purpose? & 

Can the community support the required sample size?

\vspace{.25cm}
How often have the subjects and subgroups of this work been the target of prior research?

\vspace{.25cm}
How does this work build on or differentiate itself from previously studied research questions?

\vspace{.25cm}
What community resources are being extracted vs. retained?

\vspace{.25cm}
What community resources are most important to sustaining its goals?

\vspace{.25cm}
How does the proposed work offset its footprint on the community?

\vspace{.25cm}
How will research activity be documented within the community for future researchers? \\

\bottomrule
\end{tabular}
\end{table}

\section{Pre/Post-Workshop Survey Additional Information}
\label{app:sec:community}

Below, we provide additional details about the pre and post-workshop survey questions (see Section~\ref{subsubsec:post}).
Specifically, we collected data along five key CBPR dimensions~\cite{wallerstein2020engage}.

\begin{figure}[h]
    \centering
    \begin{subfigure}{0.45\textwidth}
        \centering
        \includegraphics[width=\linewidth]{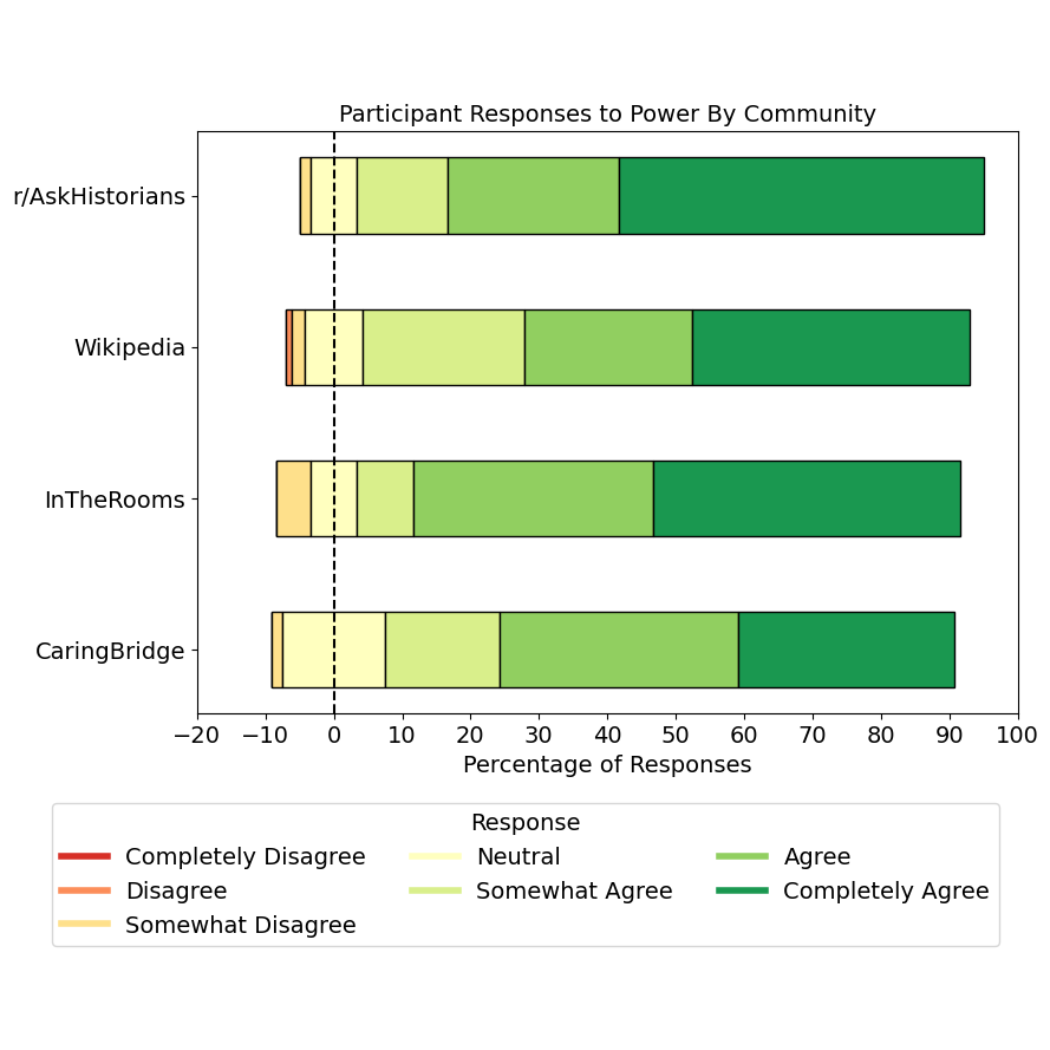}
        \caption{Power to Apply Research to the Community}
        \label{fig:image1}
    \end{subfigure}
    \begin{subfigure}{0.45\textwidth}
        \centering
        \includegraphics[width=\linewidth]{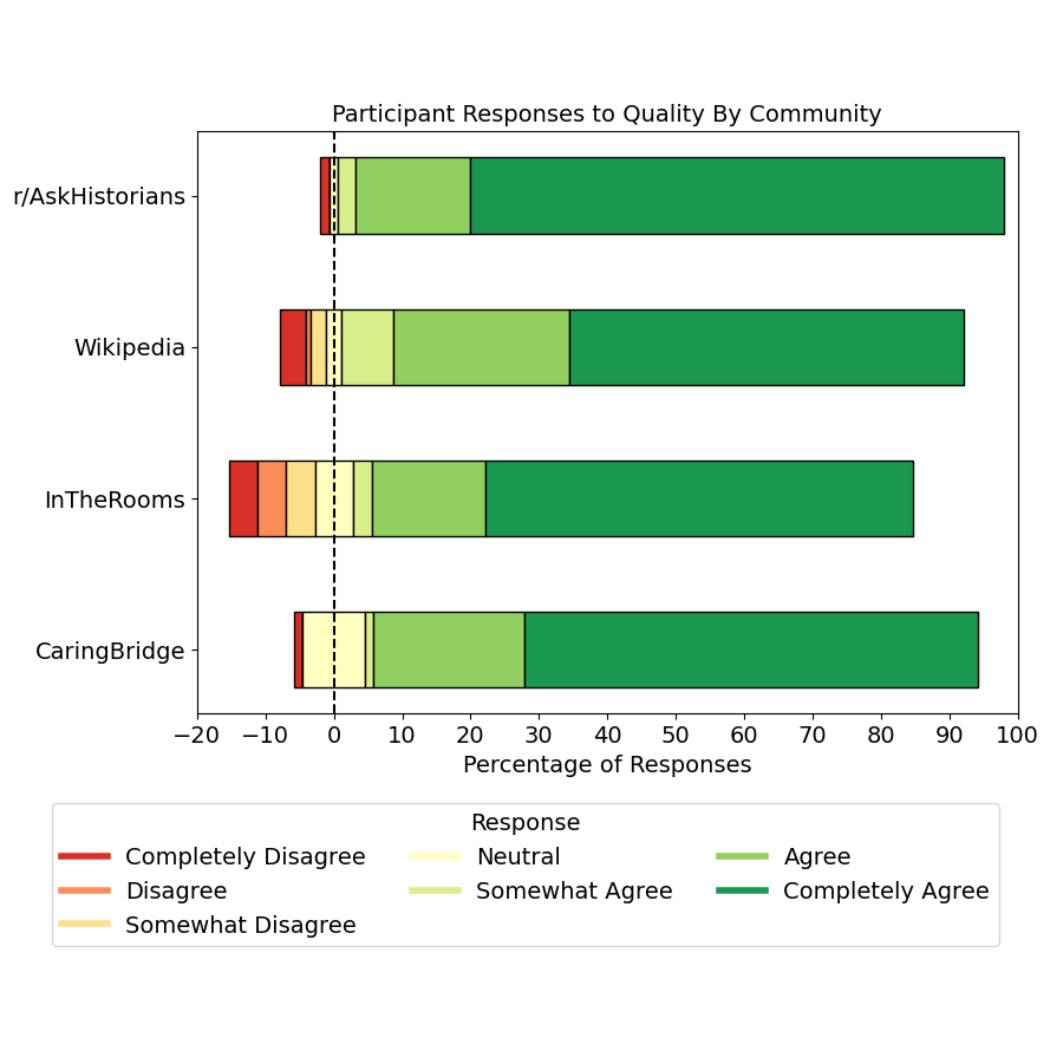}
        \caption{Influence to Impact the Research}
        \label{fig:image2}
    \end{subfigure}
    
    \begin{subfigure}{0.45\textwidth}
        \centering
        \includegraphics[width=\linewidth]{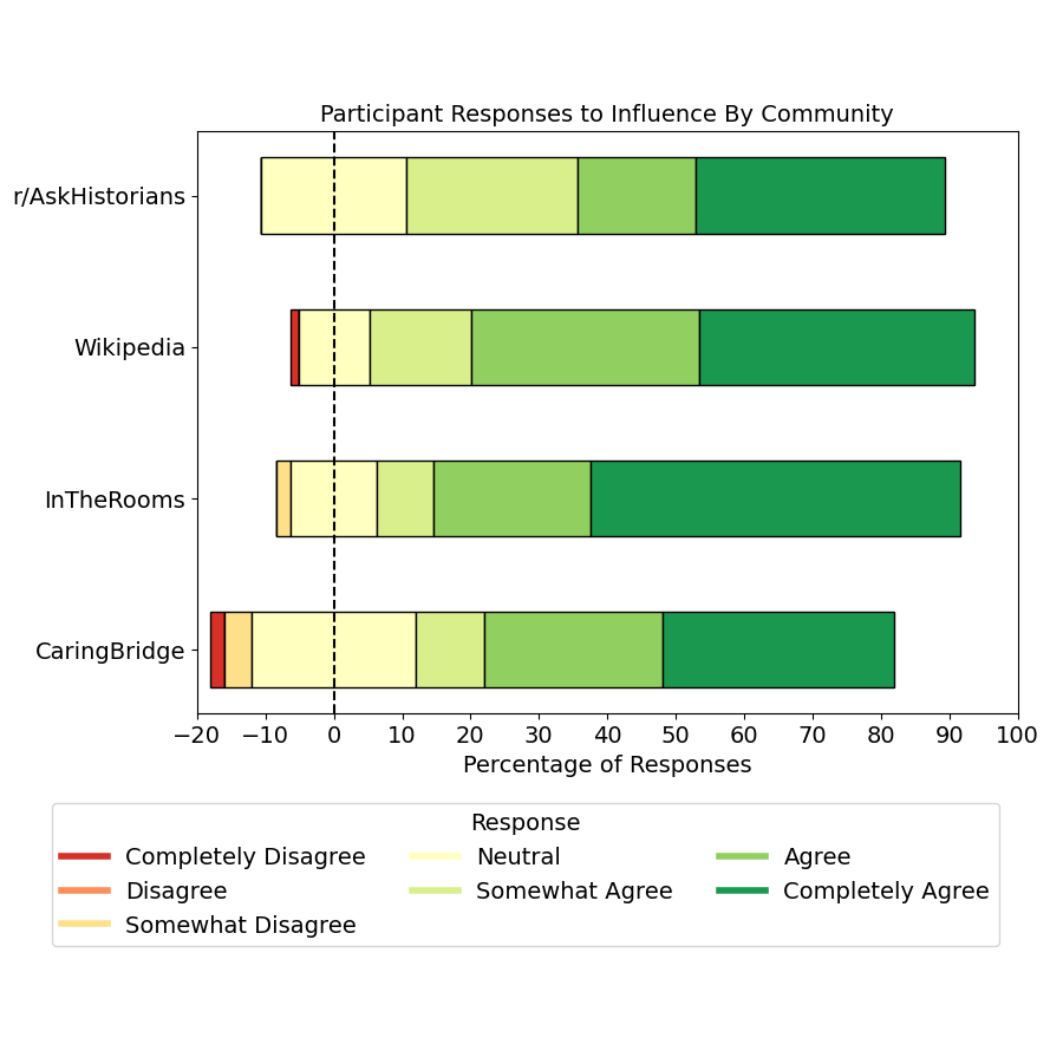}
        \caption{Quality Conversations Between Stakeholders}
        \label{fig:image3}
    \end{subfigure}
    \begin{subfigure}{0.45\textwidth}
        \centering
        \includegraphics[width=\linewidth]{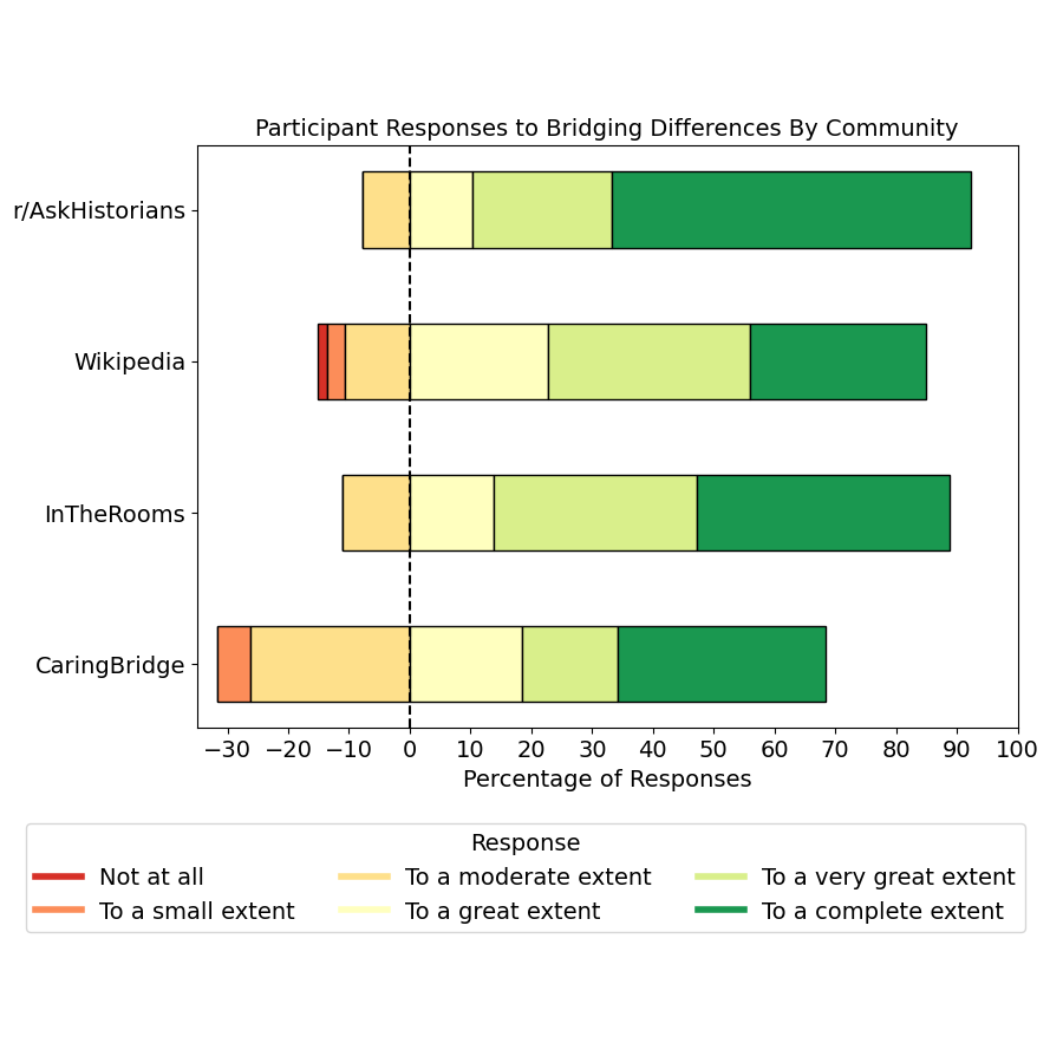}
        \caption{Bridging Differences Between Stakeholders}
        \label{fig:image4}
    \end{subfigure}
    
    \caption{Bar charts displaying the distribution of responses by community to each of the four question categories (Bridging Differences, Quality, Influence, and Power) based on the Community Engagement Survey~\cite{wallerstein2020engage}. Questions in Quality, Influence, and Power were scored on a 7-point Likert scale, reflecting the extent to which participants agreed that their experience reflected the values corresponding to each category. Questions in Bridging Differences were assessed on a Likert scale evaluating the extent to which participants felt our research team facilitated positive dialog during workshops. Overall, participants had very favorable perceptions of the [Redacted for Review]-community partnership.}
    \label{fig:2x2}
\end{figure}

One goal of post-workshop survey was to give participants the opportunity to provide feedback on their workshop experience and the research-community partnership in four main categories: Bridging Differences, Quality, Influence, and Power. Questions in Bridging Differences gauge the extent to which our research team created positive dialog with participating community partners. Questions in Quality centered around the quality of workshop discussion – namely, how much participants agree or disagree with statements such as \textit{“we show positive attitudes towards one another”} and \textit{“when conflicts occur, we work together to resolve them.”} Questions in Influence assessed participants’ perceived magnitude of impact on our research team. Questions in Power asked participants to report the extent to which they agree/disagree with statements such as \textit{“[community] members have increased participation in the research process”} and \textit{“[community] members can apply the findings of the research to practices and programs in the community”}. 
The bridging differences scale related to our ability to create positive dialog with and between participants.

Participant responses were aggregated by community and question category (see Figure~\ref{fig:2x2}). We observed that participants submitted mostly positive responses for each of the evaluation dimensions. This was important as it indicated that participants felt the workshop facilitated strong collective empowerment, community-research relationships, and community action~\cite{wallerstein2020engage}.

In addition to questions in the four aforementioned categories, participants shared their perspectives on the type of trust between their community and [Authors' Institution] during both the pre-workshop and post-workshop surveys.
Participants were asked \textit{"What primary type of trust do you think [Community] and [Authors' Institution] research team have now?"}
Table~\ref{tab:trust} details the types of trust participants could provide and the distribution of pre and post-workshop responses for each.
Before the workshop, participants had flattened distribution opinions ranging from \textit{Neutral} to \textit{Reflective}.
After the workshop, more than half of participants felt the partnership had reflective trust allowing for mistakes and differences.
We highlight that very few participating members were suspicious of the community-research relationship—a limitation we explore in Section~\ref{subsec:limitations}.

\begin{table}[]
\caption{Summary participant response ratios from pre and post-workshop trust surveys. The ratios indicate how many participants responded with that type of trust between their community and [Authors' Institution]. More participants believe the partnership had trust that allowed for mistakes and where differences can be resolved in the post-survey than the pre.}
\label{tab:trust}
\centering
\rowcolors{1}{offwhite}{}

\begin{tabular}{>{\raggedright} p{0.15\textwidth} >{\raggedright} p{0.5\textwidth} >{\raggedright} p{0.15\textwidth} >{\raggedright\arraybackslash} p{0.15\textwidth}}

\toprule
\rowcolor{white}
\textbf{Type of Trust} & \textbf{Description} & \textbf{Pre-Workshop Ratio} & \textbf{Post-Workshop Ratio} \\ \midrule
Deficit (suspicion) & Partnership members do not trust each other. & 0.02 & 0.02 \\
Neutral                   & Partners are still getting to know each other; there is neither trust nor mistrust. & 0.22 & 0.08 \\
Role-based                & Trust is based on members' title or role with limited or no direct interaction. & 0.15 & 0.05 \\
Functional                & Partners are working together for a specific purpose and time frame, but mistrust may still be present. & 0.23 & 0.23 \\
Proxy                     & Partners are trusted because someone who is trusted invited them. & 0.12 & 0.08 \\
Reflective                & Trust which allows for mistakes and where differences can be talked about and resolved. & 0.26 & 0.54 \\
\bottomrule
\end{tabular}
\end{table}

\end{document}